*Original article*
# Antibacterial and Antioxidant Activities of *Centeurea damascena* Methanolic Extract

Mohammad Jaafreh[a]; Khaled M. Khleifat[a]; Haitham Qaralleh[b]; Muhamad O. Al-limoun[a]

[a]Biology Department, Mutah University, Mutah, Karak, 61710, Jordan; [b]Department of Medical Laboratory Sciences, Mutah University, Mutah, Karak, 61710, Jordan

*Corresponding author: alkh_kha@hotmail.com

**Abstract:** The family Asteraceae include large number of *Centaurea* species which have been applied in folk medicine. One of the family Asteaceae members is the *Centaurea damascena* which authentically been tested for its antibacterial and antioxidant activity as well as its toxicity. The aims of the study were to determine the antimicrobial and antioxidant activities and toxicity of methanolic plant extracts of *Centaurea damascene*. The methanolic extracts were screened for their antibacterial activity against nine bacteria (*Staphylococcus aureus* ATCC 43300, *Bacillus subtilis* ATCC 6633, *Micrococcus luteus* ATCC 10240, and *Staphylococcus epidermidis* ATCC 12228, *Escherichia coli* ATCC 11293, *Pseudomonas aerugino* and *Klebsiella pneumoniae, Enterobacter aerogenes* ATCC 13048 and Salmonella typhi ATCC19430). The antibacterial activity was assessed by using the disc diffusion methods and the minimum inhibition concentrations (MIC) using microdilution method. The extracts from *Centaurea damascene* possessed antibacterial activity against several of the tested microorganisms. The MIC of methanol extract of *C. damascene* ranged from 60–1100 µg/mL. Free radical scavenging capacity of the *C. damascena* methanol extract was calculated by DPPH and FRAP test. DPPH radicals were scavenged with an IC50 value of 17.08 µg /ml. Antioxidant capacities obtained by the FRAP was 51.9 and expressed in mg Trolox $g^{-1}$ dry weight. The total phenolic compounds of the methanol extracts of aerial parts, as estimated by Folin–Ciocalteu reagent method, was about 460 mg GAE/ g. The phenolic contents in the extracts highly correlate with their antioxidant activity, ($R^2 = 0.976$) confirming that the antioxidant activity of this plant extracts is considerably phenolic contents-dependent.

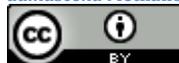

## INTRODUCTION

Medicinal plants continue being an important resource worldwide, to withstand critical diseases. Overwhelming of the world's population (60-80%) still relies on classical medicines for the curing of current disease (WHO, 2002; Zhang, 2004; Althunibat et al., 2016; Al-Asoufi et al., 2017; Majali et al., 2015; Zeidan et al., 2013; Khleifa et al., 2006; Qaralleh et al., 2009). Plants involve various compounds that have active biological substances (Mokbel and Hashinaga, 2006; Qaralleh et al., 2019). For instance, phenolic compounds and essential oil play a crucial function as vigorous natural biological factors (Cutter, 2000; Hao et al., 1998; Lis-Balchin and Deans, 1997; Puupponen-Pimia et al., 2001). The routin use of antibiotic has led to the development of one or more antibiotics resistant infectious bacteria (Sarker et al., 2005; Sufferidini et al., 2004). This issue has resulted on the failure of the treatments to numerous microbial causing infectious disease. Previous investigations pointed out to a number of medicinal plant extracts by constituting a group of potent natural antimicrobial agents (Khleifat et al., 2002; Homady et al., 2002; Rayne and Mazza, 2007; Qaralleh et al., 2009; Qaralleh et al., 2010; Tarawneh et al., 2010; Althunibat, et al., 2010; Abboud et al., 2010). Jordan's flora include more than 2500 wild plant species from 700 genera, and of these, there are around 100 endemic species, 250 rare species, and 125 quite rare species (Tellawi et al., 2001). Although, most of plant species have not been explored chemically or biologically, medicinal plants have been extensively used in traditional medication and continue being a rich source of novel therapeutic agents (Cragg et al., 1997).

Plant family Asteraceae, consists of more than 1500 genera and approximately 2500 species worldwide (Wagstaff and Breitwieser, 2002). The genus Centaurea. is a polymorphous genus belonging to the Asteraceae family, and comprises 400–700 species of annual, biennial and perennial grassy plants, rarely dwarf shrubs predominantly distributed in Europe and Asia (Dittrich, 1977; Wagenitz and Hellwig, 1996; Bancheva and Greilhuber, 2006).

The genus *Centaurea* comprises more than 500 species that are common worldwide, particularly the Mediterranean and western Asia area (Mabberley, 1997). Some Asteraceae species have been used in many fields, including nutrition and medicinal industries (Tepe et al., 2006; Qaralleh et al., 2009; Tekeli et al., 2011). Many species of Centaurea have long been utilized in conventional medication to treat different diseases, like hemorrhoids, abscess and the common cold (Kargıoglu et al., 2008; Kargıoglu et al., 2010; Sezik et al., 2001). Thus, it is worthwhile to elaborate on the antibacterial activities of different species of Centaurea on the case-by-case basis. However, most of plant species have not been explored chemically or biologically; medicinal plants have been extensively used in traditional medication and continue being a rich source of novel therapeutic agents (Cragg et al., 1997). In the present study, we demonstrate that methanolic extracts of *C damascena* which are used to treat gastritis and for flavoring hot tea on a small scale in a town south of Jordan (Bsaira). The choice of this species was based on a review of the local folk literature which points out the therapeutic properties that the genus Centaurea possess (Khleifat et al., 2007). and on a preliminary investigations conducted in our laboratory using different medicinal plants for their low cytotoxicity (Qaralleh et al., 2009; Zeidan et al., 2013; Majali et al., 2015; Althunibat et al., 2016; Al-Asoufi et al., 2017). The aim of this study was to determine the antibacterial and antioxidant activities as well as toxicity of *Centaurea damascene* that authentically been explored in this work as one of species of *Centaurea* growing in Jordan.

**MATERIALS AND METHODS**
**Plant materials**
Samples of *Centaurea damasscena* were collected in May and June from Dhana Natural Reserve (DNR), South of Jordan in June 2017 as previously reported (Khleifat et al., 2019). The plant was recognized by Dr. Salih Quraan, Department of Biology, Faculty of Science, Mutah University, Jordan. Voucher specimens were deposited in the department of biology, Faculty of Science, Mutah University, Jordan.

**Preparation of methanolic extracts**
Air-dried plant material was finely powdered using a laboratory mill. 50 g of each sample was exhaustively extracted with 500 ml methanol at room temperature under stirring and the extracts were filtered through a Whatman blue filter. After evaporation of the solvent at 40°C in rotary evaporator, the residues were stored at 4°C until further analysis.

**Microorganisms**
All bacteria were maintained at −20 °C in nutrient agar (NA) and all yeast, in Sabouraud dextrose agar (SDA, Difco) containing 20% (v/v) glycerol. Before testing, the bacteria were transferred to nutrient broth (NB) (Khleifat et al., 2003; Khleifat, 2006a-c) and the fungi, to Sabouraud dextrose broth (SDB, Difco), and were cultured overnight at 37 °C. Then, the turbidity was adjusted to equivalent to 0.5 McFarland standards (approximately $10^8$ CFU/mL for bacteria and $10^5$ or $10^6$ CFU/mL for fungi). The antibacterial activities of the *Centaurea damascena* were assessed against nine bacterial species. The test organisms included four Gram-positive bacteria [*Staphylococcus aureus* ATCC 43300, - *Bacillus subtilis* ATCC 6633, *Micrococcus luteus* ATCC 10240, and *Staphylococcus epidermidis* ATCC 12228]; five Gram-negative bacteria (*Escherichia coli* ATCC 11293, *Pseudomonas aeruginosa* and *Klebsiella pneumoniae, Enterobacter aerogenes* ATCC 13048 and *Salmonella typhi* ATCC19430). They were all obtained as pure cultures from the Department of Biology, Faculty of Science, Mutah University, Mutah, Jordan.

**Evaluation of antimicrobial activity**
The antimicrobial activity of each methanol extract sample was evaluated by using the disc-diffusion method. The spreading method performed according to Klančnik et al., (2010), 100μl of each bacterial suspension was uniformly spread on a solid growth medium in a Petri dish. Sterile Whatmann No. 1 filter paper discs (6 mm) were soaked with 25 μL of extract residue diluted in the corresponding extract solvent (1000 μg/1 mL of 12.5% dimethyl sulfoxide (DMSO)) and placed on the surface of the freshly inoculated medium. The media were incubated for 24 h at 37 °C. Antibiotic susceptibility discs including bacitracin (0.04 U), ceftazidime (30 μg), imipenem (10 μg), novobiocin (5 μg), polymyxin B (300 U), tetracycline (30 μg), ampicillin (10 μg) and cycloheximide were used as control, and negative controls were 12.5% DMSO, methanol and deionized water. The antimicrobial activity was evaluated by measuring the diameter of the inhibition zones (Khleifat et al., 2006a; Khleifat et al., 2006b; Khleifat, 2006a; Tarawneh et al., 2009). The experiment was performed in triplicate. The values were expressed as means with standard deviations (±SD).

**Microdilution method**
Minimum inhibition concentration (MIC) values were determined using 96-well microtiter plates by dissolving the sample in DMSO. Serial dilutions were made to obtain concentrations ranging from 3 to 0.0137 mg/ml. Suspensions of standard microorganisms were inoculated onto the microplates. The growth of the microorganisms was observed by using a microplate photometer (Thermo Scientific Multiskan). The MIC values were defined as the lowest concentrations of the plant extracts to inhibit the growth of microorganisms.

**Determination of free-radical-scavenging activity DPPH (1,1-diphenyl-2-picrylhydrazyl) method**

The free-radical-scavenging activity assay of the plant extracts was measured by using DPPH (1,1-diphenyl-2-picrylhydrazyl) method (Blois, 1958; Kumar et al., 2011). The plant sample powder was dissolved in ethanol giving a concentration of 10 mg/mL as stock solutions. Different concentrations of solutions (30, 25, 20, 10 and 5 µg/mL) were made via the serial dilution methods. Four mL of a DPPH-ethanol solution (0.1 mmol/L) was mixed with one millilitre of ethanol extract solution for each concentration. After thirty minutes, the absorbance was determined at 517 nm (UV-VIS Spectrophotometer UV-9200). The activity of DPPH-radical-scavenging ability was calculated by using the equation:

$$\% \text{ inhibition} = [(AB - AS)/AB] \times 100\%$$

where AB is the absorbance of the positive control and AS is the absorbance of the sample containing the tested extract. Ascorbic acid was used as a standard or positive control. Reaction mixture without a sample was used as the negative control (Faruk et al., 2016). All experiments were run in triplicate. The concentration providing 50% inhibition ($IC_{50}$) was calculated from the resulting graph plot by interpolation.

**Ferric reducing antioxidant power (FRAP) method**

The FRAP assay was carried out according to the method described by Benzie and Strain, (1996). The FRAP assay was based on the reducing power of antioxidants in which a potential antioxidant will reduce the ferric ions to the ferrous ions, which form a blue colored ferrous-tripyridyltriazine complex. The FRAP reagent was freshly prepared by mixing together 10 mM 2,4,6-tripyridyl triazine (TPTZ) (1 mL) and 20 mM ferric chloride (1 mL) in 0.25 M acetate buffer (10 mL, pH 3.6). Plant extract sample (50 µL) was added to 3 mL of the FRAP reagent (The final concentration of the plant extract in the solution was 100 µg/mL). The tests were carried out in triplicate. The absorbance was measured at 593 nm after 8 min incubation at room temperature. Trolox (final concentration 0 to 1 µg/mL) was used as a standard for the construction of the calibration curve. The antioxidant capacity based on the ability to reduce ferric ions of the extract was expressed as mg Trolox® equivalents per g of plant extract.

**Phytochemical analysis**

The phytochemical screening were carried out for the plant extract using the standard procedures (Kokate, 2005).

**Determination of total phenolic content (TPC)**

The content of total phenolic compounds in plant methanolic extracts was determined by Folin–Ciocalteu method according to Miliauskasa et al., (2004), 0.2 ml of the plant extract (0.5 mg / ml) was mixed with 1 ml of 10% Folin–Ciocalteu solution and 0.8 ml of 7.5% sodium carbonate solution. The mixture was incubated for 1 h at room temperature. The absorbance at 760 nm was measured and converted to phenolic contents according to the calibration curve of gallic acid. The Gallic acid curve equation were taken from Miliauskasa, Venskutonisa and Beek, 2004. All determinations were performed in triplicate. Total content of phenolic compounds in plant methanol extracts in gallic acid equivalents (GAE) was calculated by the following formula: C= c.V/m Where: C-total content of phenolic compounds, mg/g plant extract, in GAE; c-the concentration of Gallic established from the calibration curve, mg/ml; V- the volume of extract, ml; m-the weight of pure plant methanolic extract, g

**RESULTS**
**Antimicrobial Activity**

From aerial part of whole plant of C. damascene, the average yield of ethanol extract obtained was 11.44% (v/ w) on dry weight basis (Table 1). The antimicrobial activities of methanolic extract of C. damascene was tested in vitro by using the disc diffusion method and minimal inhibitory concentration (MIC) using microdilution assay. The *C. damascena* methanolic extracts showed varied antimicrobial activity against the bacteria used in this study (Tables 2 and 3). Among the tested microorganisms, *M. lutes* was the most susceptible microorganism against lowest concentration of methanolic extracts (500µg/ml). The *C. damascena* extract (2000µg/ml) showed good inhibitory activity on most of the bacterial genera tested. The methanol extract showed highest antibacterial activity against *E. aerogenes and S. epidermidis* (19.3 mm) followed by *S. aureus* (18 mm), *B. suubtilis* and *M. lutes* (17 mm) (Table 2). For a more reliable assessment of antimicrobial activity, a broth dilution assay was carried out. The susceptibility of the test microorganisms against active extracts was evaluated and results are shown as MIC (Table 3). The lowest MIC value was recorded by *C. damascena* against *M. lutes* (15 µg mL/l) in consistency with the result of disc diffusion method. MIC values of *C. damascena* against *E. coli*, *P. aruginosa, K. pneumonia, S. typhi* were almost 1100 µg/ml and for *E. aerogenes* was 123 µg/ml. The extracts from *Centaurea damascene* possessed antibacterial activity against several of the tested microorganisms. The MIC of methanol extract of *C. damascena* ranged from 60-1100 µg/mL.

Table 1. Yield of ethanol crude plant extracts (%)

| Plants | Dry weight (g) | Weight of ethanol extract (g) | Yield of ethanol extract (%) |
|---|---|---|---|
| *C.damascena Bioss* | 50 | 5.72 | 11.44% |

Table 2. Antibacterial activity of methanol extract of *C. damascene* (Zone of inhibition expressed as millimeter (mm)). Data are expressed as mean ± SD, where n = 3..

| Bacterial strains | Zone of inhibition (mm) | | | | |
|---|---|---|---|---|---|
| | 500 µg/disc Mean ± SD | 1000µg/disc Mean ± SD | 2000µg/disc Mean ± SD | Ampicillin | Cefixime |
| *B. subtilis* | 12 ± 0.62 | 15 ± 0.42 | 17 ± 0.50 | 10 ±0.30 | 20 ±0.30 |
| *E. aerogenes* | 9 ±0.50 | 13 ± 0.45 | 19 ± 0.53 | - | 25 ±0.20 |
| *E. coli* | - | - | - | 10 ±0.20 | 23 ±0.15 |
| *S. epidermidis* | 9 ±0.30 | 13 ±0.6 | 19 | - | 10 ±0.33 |
| *M. luteus* | 12 ±0.30 | 15 ±0.20 | 17 ±0.70 | - | 9 ±0.20 |
| *K. pneumoniae* | - | - | - | - | 20 ±0.30 |
| *P. aeruginosa* | - | - | - | - | 23 ±0.20 |
| *S. Typhi* | - | - | - | - | 36 ±0.20 |
| *S. aureus* | 9 ±0.56 | 13 ±0.40 | 18 ±0.30 | 15 ±0.40 | 15 ±0.15 |

(-), resistant strains

Table 3. Antibacterial activity of methanolic extract of *C. damascena* (MIC values expressed as µg/ml).

| Bacterial strains | Methanol extract |
|---|---|
| *B. subtilis* | 123 |
| *E. aerogenes* | 123 |
| *E. coli* | 60 |
| *S. epidermidis* | 1100 |
| *M. luteus* | 1100 |
| *K. pneumoniae* | 1100 |
| *P. aeruginosa* | 121 |
| *S. Typhi* | 1000 |
| *S. aureus* | 123 |

Table 4. Antioxidant activity and total phenolic content of methanolic extracts of *C. damascena* from Jordan.

| Plant Scientific name | TEAC (mg/gdw)[a] | | | |
|---|---|---|---|---|
| *C. damascena* | FRAP | DPPH | IC50 (µg/ml)[c] | Total phenolic[b] |
| | 51.9±1.0 | 43±0.75 | 17.08±0.55 | 460±3.75 |

[a]TEAC, Trolox Equivalent Antioxidant Capacity (TEAC) (mg Trolox $g^{-1}$ dry weight). Absorbance was converted to equivalent activity of Trolox per g of dry weight based on a standard curve. [b]Total phenolic content expressed as mg of gallic acid equivalent (GAE)/g of dry weight (dw)., Results of DPPH assay mean of triplicate assays,±SD;

## Antioxidant Activity

Free radical scavenging capacity of the *C. damascna* methanol extract was calculated by DPPH test (Table 4). IC50 value is expressed by which the functional concentration of 50% DPPH radicals were scavenged and was determined from the graph plotting inhibition percentage versus extract concentration. $IC_{50}$ of methanol extract was detected by using DPPH assay and expressed as µg/ml. *C. damascena* extract exerted antioxidant efficiency with an IC50 value of 17.08 µg /ml. The extract exhibited concentration-dependent radical scavenging activity.

Antioxidant capacities obtained by the FRAP was 51.9 and expressed in mg Trolox $g^{-1}$ dry weight. The total phenolic compounds of the methanol extracts of aerial parts, as estimated by Folin–Ciocalteu reagent method, was about 460 mg GAE/g (Table 5). The phenolic contents in the extracts highly correlate with their antioxidant activity, ($R^2 = 0.98$) (Fig. 1), proving that phenolic compounds involved considerably to the antioxidant activity of these plant extracts.

Table 5. Qualitative analysis of phytochemicals in methanol extract of *C. damascena*

| Phytochemical | *C. damascena* |
|---|---|
| Tannins | + |
| Terpenoids | + |
| Saponins | + |
| Glycosides | + |
| Flavonoids | + |
| Steroids | - |
| Alkaloids | + |

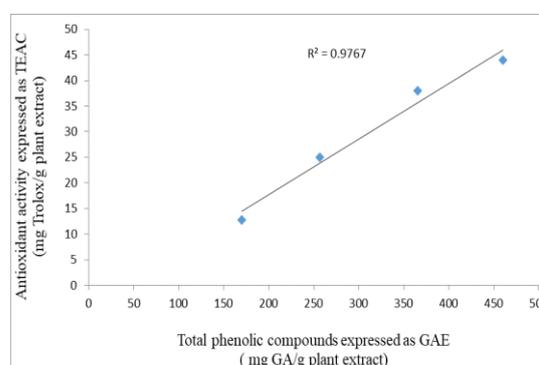

Figure 1. Correlation between TPC and antioxidant activity obtained by DPPH

## DISCUSSION

The regularly used method for evaluation of antibacterial efficiency of publically used medicinal plants can be diffusion as qualitative and dilution methods of the extracts as quantitative analyses (Vanden Berghe and Vlietinck, 1991; Köse et al., 2008). The antibacterial activities of the methanolic extracts of *Centaurea damascena* were assessed by using disc diffusion methods and the MIC values via a microdilution method. Essential oil constituents were previously isolated from C. damascena (Khleifat et al., 2019). It was reported that the essential oil components including Fokienol, thymol, Alpha Terpineol, Chrysanthemumic acid Terpinen-4-ol and Borneol were prevailing with a high degree of polymorphism in the incidence of these compounds between different species of *Centaurea* (Khleifat et al., 2019).

The results show that plant extracts have antibacterial activity, although plant extracts did not have that effect as essential oils do on all the tested bacteria (Khleifat et al., 2019). The antimicrobial activity of other *Centaurea* species other than *damascena* was focused by many authors (Hajjeh et al., 1999; Rubin et al., 1999; Willet, 1992; Hillard et al., 2003; Güven et al., 2005; Sarker et al., 2005; Buruk et al., 2006; Köse et al., 2007; Yayli et al., 2009; Ugur et al., 2010; Cansaran et al., 2010 and Tekeli et al., 2011). The results gotten from the disc diffusion tests showed that there has been an increasing action of plant extracts on the inhibition of bacterial growth on concentration dependent manner. The observed activity may be due to the presence of potent phytoconstituents in the plant extract. This may be indicative of a significant potential for isolating purer compounds. The antibacterial activity of most examined *Centaurea* species were shown to be weak or solvent-dependent; for example, the previously examined *Centaurea* species showed weaker antibacterial activity than the ethyl acetate extract of *C. cankiriense* (Sarker et al., 2005). For example, when employing microdilution assay the *C. hubermorathii, C. bornmuelleri,* and *C. schiskinii* showed no any activity on *E. coli* and *B. cereus* (Tekeli et al., 2011).

However, it has no data previously been reported to discuss any possible varying degree of antibacterial activity by whole *C. damascena* against any type of microorganisms. Tekeli et al., (2011), conducted a study on the antimicrobial and antioxidant activities of 12 different species of *Centaurea* as Turkish flora. Two of the 12 studied *Centaurea species with 4 mg/ml concentration only inhibited B. cereus*. In Tekeli and his team's study, *C. calolepis* displayed an action on *Salmonella enteritidis*, while *C. urvillei subsp. urvillei* was active against *Escherichia coli* and *Staphylococcus aureus*. In further study on *C. cankiriense* the antimicrobial activity of ethylacetate and methanol extracts displayed growth inhibitory effect against 13 bacteria with MIC values of the ethyl acetate extracts were being as 250 and 62.5 µg/ml for *E. coli* and *S. aureus*, respectively (Cansaran et al., 2010).

Güven et al. (2005) reported an important antimicrobial effect of five *Centaurea* species (*C. ptosomipappoides, C. odyssei, C. ptosomipappa, C. amonicola* and *C. kurdica*) on *S. aureus* and *B. cereus*. Conversely, extracts of *C. appendicigera* and *C. helenioides* did not display antimicrobial activity on *E. coli* (Buruk et al., 2006). The chloroform and ethyl alcohol extracts of *C. cariensis* subsp. *Niveotomentosa* exhibited powerful antibacterial activities on various resistant bacteria, particularly *Staphylococcus* strains (Ugur et al., 2010).

The high antibacterial efficacy in the methanolic extract possibly be due to the presence of tannins, flavonoids, and terpenoids (Zablotowicz et al., 1996; Hemandez et al., 2000; Mamtha et al. 2004; Cowan, et al., 1999 ). These medically bioactive ingredients practice antimicrobial activity through various mechanisms. Flavonoids, which have been found to be effective antimicrobial substances against a wide array of microorganisms in vitro, are known to be synthesized in response to microbial infection by plants. They have the ability to complex with extracellular and soluble proteins and to complex with bacterial cell walls (Cowan, et al., 1999). The saponins have the capacity to rise leakage of metabolites from the cell (Zablotowicz et al., 1996). Tannins cause cell wall synthesis inhibition by forming irreversible complexes with prolene rich protein (Mamtha et al. 2004). Terpenoids cause dissolution of the cell wall of microorganism by undermining the membranous tissue (Hemandez et al., 2000). Furthermore, plant extracts are lacking steroids and this probably why Gram-negative bacteria were less susceptiple to *C. damascena* extract than the Gram-positive one. The steroids are known for their antibacterial activity particularly connected with membrane lipids and cause leakage from liposomes (Epand et al., 2007).

DPPH is normally employed as a reagent to estimate free radical scavenging capacity of antioxidants (Oyaizu, 1986; Zengin et al., 2010). Free radical scavenging capacity of the *C. damascna* methanol extract was calculated by DPPH test (Table 1). IC50 value is expressed by which the functional concentration of 50% DPPH radicals were scavenged and was determined from the graph plotting inhibition percentage versus extract concentration. $IC_{50}$ of methanol extract was detected by using DPPH assay and expressed as µg/ml. IC50 estimate is inversely correlated to antioxidant capacity of extracts. In the present study, *C. damascena* extract exerted better antioxidant efficiency than that of previously reported *Centaurea* species extract with an IC50 value of 17.08 µg /ml. 25.12e54.14% and 27.03e63.19%, respectively. The extract exhibited concentration-dependent radical scavenging activity, As far as we know from literature survey, the *Centaurea damascena* have shown better activities than other *Centaurea* species such as *C. patula, C. pulchella,* and *C. tchihatchfeii* (Zengin et al., 2010), *C. mucronifera* (Tepe et al., 2006), *C. huber-morathii* (Sarker et al., 2005) and *C. centaurium* (Conforti et al., 2008) by the same DPPH assay. Antioxidant capacities obtained by the FRAP was 51.9 and expressed in mg Trolox $g^{-1}$ dry weight. The two methods (FRAP and DPPH) approximately showed similar results for *C. damascena* methanol extract since the two methods, the assay results were generally similar for different plant models because these methods are based on electron transfer mechanism (OZGEN et al., 2006)

The total phenolic compounds of the methanol extracts of aerial parts, as estimated by Folin–Ciocalteu reagent method, was about 460 mg GAE/ g. The highest level of phenolics was found in *C. damascena* as compared with previously reported other species of *Centuarea* (Sarker et al., 2005; Tepe et al., 2006; Karamenderes et al., 2007; Alali et al., 2007; Conforti et al., 2008; Dudonné et al., 2009; Zengin et al., 2010). The results of total phenolic compounds suggested that the phenolic contens contributed significantly to the antioxidant ability of the *C. damascena extract*. It has been reported in the literature that phenolic contents have strong antioxidant capacities and these phenolics have antioxidant activity mainly because of their redox properties, which enable them to act as reducing agents, singlet oxygen quenchers, hydrogen donators, and metal chelator (Rice-Evans et al., 1995; Baratto et al., 2003; Alali et al., 2007; Zengin et al., 2010). However, it was found that methanolic plant extracts are the most effective scavenger of DPPH radical as compared with other solvents (Miliauskasa *et al*., 2004). Hence, it was suggested that methanol is more efficient solvent for cell walls and seeds degradation, that have nonpolar nature causing the release of polyphenols from cells. It can be observed that the phenolic contents in the extracts highly correlate with their antioxidant activity, ($R^2$ = 0.87), confirming that phenolic compounds contribute significantly to the antioxidant activity of these plant extracts. The enormous disparity in *Centaurea* species antioxidant activity could result from variations in total phenolic content**s** (Al-Mustafa, & Al-Thunibat, 2008). Such observation agreed with several previous findings (Sarker et al., 2005; Tepe et al., 2006; Karamenderes et al., 2007; Alali et al., 2007; Conforti et al., 2008; Zengin et al., 2010). Moreover, phenolic compounds have various reactions to Folin-Ciocalteau assay (Sun et al., 2002). The molar response of this method is roughly proportional to the number of phenolic hydroxyl groups in a given substrate, but the reducing capacity is enhanced when two phenolic hydroxyl groups are oriented in ortho or para-position. Since these structural features of phenolic compounds are responsible for antioxidant activity (Katalinic *et al*., 2006). Thus, polyphenols measurements in extracts may be related to their antioxidant activities.DPPH and FRAP methods have been used by many researchers to evaluate the free radical scavenging activity of antioxidant molecules and plant extracts. DPPH does generate strongly colored solutions with methanol which is eliminated in presence of antioxidants. The data obtained from the two radical scavenging methods suggest high accuracy and constancy between them. Using two methods, some plants showed large difference in their TEAC values, whereas others showed little differences. This may be due to variation in types of phenolic compounds, that differ significantly in their reactivity towards DPPH (Katalinic *et al*., 2006). Furthermore, the affinities of different Centaurea species toward the two above radicals were sometimes significantly altered due to variation of DPPH solubility in aqueous medium. However, the role of optimized growth conditions have been strongly referred to as a key factor in the activity of various enzymes and this is reflected in the physiological response of many bacteria (Khleifat et al., 2006b; Khleifat et al., 2015; Al-Limoun et al., 2019).

**CONCLUSION**
Finally, it must be said that always-growing conditions are also an influential factor in responding to many bacterial activities particularly physiological features (Khleifat, 2007; Khleifat, 2006b; Khleifat, 2006c; Khleifat, 2006d; Shawabkeh et al., 2007). The maximum zone of inhibition was observed at the concentration of essential oil and methanol extract of 10 µg/disc and 2000 µg/disc, respectively. Moreover, purification and fractionation will clarify the possible compound, which is an imperative requirement, because of the raising in resistance of the presently available antibiotics. The present study enhance the existing information of presence of different phytochemical active compounds in *C. damascena* possessing significant broad-spectrum antibacterial effectiveness.